\documentclass[
onecolumn,
amsmath,amssymb,
nofootinbib,
showkeys,
prd,
aps,
10pt,
longbibliography,preprintnumbers
]{revtex4-2}
\pdfoutput=1
\usepackage{hyperref}
\usepackage{xcolor,graphicx}% Include figure files
\usepackage{bm}% bold math

\begin{document}

\title{Gravity from thermodynamics in vacuum: Lorentz invariance and the role of Bel-Robinson tensor}

\author{Ana Alonso-Serrano}
\affiliation{Institut für Physik, Humboldt-Universität zu Berlin, Zum Großen Windkanal 6, 12489 Berlin, Germany, and}
\affiliation{Max-Planck-Institut f\"ur Gravitationsphysik (Albert-Einstein-Institut), \\Am M\"{u}hlenberg 1, 14476 Potsdam, Germany}
\email{ana.alonso.serrano@aei.mpg.de}

\author{Marek Li\v{s}ka\footnote{Corresponding author}}
\affiliation{School of Theoretical Physics, Dublin Institute for Advanced Studies, 10 Burlington Road, Dublin 4, Ireland, and}
\affiliation{Institute of Theoretical Physics, Faculty of Mathematics and Physics, Charles University, V Hole\v{s}ovi\v{c}k\'{a}ch 2, 180 00 Prague 8, Czech Republic}
\email{liskama4@stp.dias.ie}

\vspace{10pt}

\begin{abstract}
Thermodynamics of local causal horizons has been shown to encode the information necessary to derive the equations governing the gravitational dynamics. We have previously argued that, in the presence of matter, this derivation further implies quantum phenomenological corrections to gravitational dynamics. Herein, we study whether similar corrections also occur in vacuum. We show that, under the assumptions of locality and local Lorentz invariance of physics, the vacuum dynamics is prescribed by the Einstein equations. We also discuss an alternative paradigm which assumes the existence of a preferred direction of time (much like in Einstein-aether or Ho\v{r}ava-Lifshitz gravity). Then, we find an modified gravitational dynamics in which Ricci curvature is sourced by Bel-Robinson super-energy.
\end{abstract}

%\keywords{Bel-Robinson tensor}

\maketitle

\section{Introduction}

Gravitational dynamics shares a strong connection with thermodynamics~\cite{Bekenstein:1973,Bardeen:1973,Wald:1993,Jacobson:1995ab,Curiel:2014}. This relation becomes especially clear in the way the heat term appears in the first law of black hole thermodynamics. This term can be derived directly from the gravitational Lagrangian as a conserved charge associated with the time translational Killing symmetry~\cite{Wald:1993,Wald:1994,Iyer:1996,Compere:2018,Margalef:2021,Hollands:2024}, without imposing any further assumptions. Then, quantum field theory in curved backgrounds shows that black holes emit thermal radiation, assigning a physical temperature to the black hole~\cite{Hawking:1975} (the Hawking temperature). With these two ingredients, the ratio of the heat term in the first law of black hole thermodynamics and the Hawking temperature yields the expression for the black hole entropy. Notably, no similarly straightforward link between the dynamics and the entropy appears in other physical theories, making gravity unique in this regard.

The black hole entropy not only follows directly from the gravitational Lagrangian, but actually contains enough information to reconstruct the gravitational dynamics~\cite{Jacobson:1995ab,Eling:2006,Padmanabhan:2008,Padmanabhan:2010,Chirco:2010,Jacobson:2012,Faulkner:2014,Faulkner:2017,Jacobson:2015,Bueno:2017,Svesko:2018,Svesko:2019,Alonso:2020,Alonso:2021,Alonso:2024}. It requires extending the gravitational entropy from being specific to black holes to a universal property of any causal horizon. This requirement is supported by the analysis of thermodynamics of various horizons, including de Sitter, cosmological and acceleration horizons~\cite{Gibbons:1977,Sorkin:1986,Srednicki:1993,Jacobson:2003,Solodukhin:2011,Jacobson:2023,Jacobson:2023b,Alonso:2024}. In each of these case, the entropy takes the same form as for the black hole horizon. Then, constructing a local causal horizon in every regular spacetime point, we can analyse the behaviour of their entropy. It turns out that the thermodynamic equilibrium conditions imposed on such local horizons encode (locally) the equations governing the gravitational dynamics. If the equivalence principle\footnote{More precisely, one in general needs the Einstein equivalence principle. Specifically, to recover the Einstein equations a more stringent formulation, the strong equivalence principle (valid for gravitational test physics) is required~\cite{Chirco:2010}. For detailed discussion of the various formulations of the equivalence principle and their relations, see~\cite{Casola:2015}.} holds, these equations are valid throughout the spacetime. For entropy proportional to the horizon area, this approach recovers Einstein equations~\cite{Jacobson:1995ab,Chirco:2010,Jacobson:2015}. If the input of the derivation is any Wald entropy density~\cite{Wald:1994} constructed from the metric and the Riemann tensor, it yields the equations of motion of the corresponding modified theory of gravity~\cite{Jacobson:2012,Bueno:2017,Svesko:2018,Svesko:2019}. Moreover, one can phrase the local equilibrium conditions entirely in terms of quantum von Neumann entropy. Then, they encode the semiclassical Einstein equations in which the quantum expectation value of the energy-momentum tensor serves as the source term for the classical spacetime curvature~\cite{Faulkner:2014,Faulkner:2017,Jacobson:2015,Bueno:2017}.

Herein, inspired by this success of thermodynamic methods in recovering the classical and semiclassical gravitational dynamics, we employ them to peer one step further. Namely, we study the thermodynamic perspective on the low energy quantum gravitational effects. We follow a research program we started some years ago, looking at the leading order quantum gravitational correction to entropy of causal horizon. It turns out to be proportional to the logarithm of the horizon area. Remarkably, a leading order correction term of this form is nearly universally predicted by various approaches to quantum gravity~\cite{Kaul:2000,Meissner:2004,Banerjee:2011,Sen:2013,Karan:2023,Faulkner:2013} and also by some theory-independent considerations~\cite{Mann:1998,Solodukhin:2010,Adler:2001,Medved:2004,Medved:2005,Das:2002,Gour:2003,Hod:2004,Davidson:2019}. This generality of the logarithmic term in horizon entropy makes its influence on gravity robust and independent of any specific approach to quantum gravity. We have previously studied the effect of this term on the linearised gravitational dynamics. In this case, we found a result equivalent to the linearised equations of motion of quadratic gravity. We have also shown that beyond the linearised regime, the logarithmic contribution to entropy implies correction terms quadratic in the Ricci tensor. Notably, such terms do not appear in quadratic gravity, marking a clear departure from it. However, in both special cases we studied so far the corrections to gravitational dynamics only occur in the presence of matter. In the present work, we ask whether the Einstein equations are modified even in vacuum. We show that imposing the local Lorentz invariance prohibits any such modifications.

A more speculative option lies in breaking the local Lorentz invariance by introducing a preferred local direction of time. Then, the thermodynamic derivation yields a single scalar equation resembling the Hamiltonian constrain of general relativity. The source term of this equation is proportional to the Bel-Robinson tensor contracted with the local direction of time in all indices~\cite{Bel:1958,Senovilla:2000}. Such a scalar expression is known as the Bel-Robinson super-energy density. The Bel-Robinson tensor is quadratic in the Weyl tensor and satisfies an analogy of the dominant energy condition valid for the energy-momentum tensor. It has been suggested that the Bel-Robinson super-energy density can be interpreted as a quasilocal gravitational energy density per unit surface area~\cite{Garecki:1977,Senovilla:2000}. This interpretation is consistent with the role the Bel-Robinson super-energy density plays in the geometry of small spheres, where it accounts for the area deficit due to spacetime curvature in vacuum~\cite{Horowitz:1982,Bergqvist:1994}. Then, its appearance as a source term for the Hamiltonian constraint is somewhat intuitive, as in general relativity the source is the matter-energy density. Moreover, it is well established that both the Bel-Robinson super-energy~\cite{Horowitz:1982,Bergqvist:1994,Jacobson:2017} and other measures of the gravitational energy~\cite{Torre:1986,Brown:1993b,Feng:2022} play a role in the dynamics of surfaces and, hence, in the gravitational thermodynamics. It is then to be expected that they can also act as source terms in the thermodynamically derived equations for gravitational dynamics. In total, while we leave the in depth exploration of this Lorentz invariance violating setup for future works, we find the possibility interesting enough to warrant a discussion in the present paper.

In this work, we realise the local causal horizons as causal diamonds. The changes of the area of the horizon of causal diamonds in vacuum and their relation to the Bel-Robinson tensor have been explored in previous works~\cite{Jacobson:2017,Wang:2019}. We build on the results of these references, connecting them with the derivation of the equations governing the gravitational dynamics. This connection is made possible by two innovations we introduce compared to the previous works. First, we work with the light-cone cut construction of the local causal diamonds, slightly modifying its original version~\cite{Wang:2019}, to make it suitable for deriving the gravitational equations. Second, as we discuss in section~\ref{Weyl corrections}, considering the logarithmic correction to entropy allows us to obtain nontrivial local corrections to the traceless Einstein equations.

The paper is organised as follows. In section~\ref{Einstein}, we briefly recall the derivation of the traceless Einstein equations from thermodynamics of local causal diamonds. Section~\ref{Weyl corrections} derives the corrections in vacuum, finding them to be proportional to the Bel-Robinson tensor in four spacetime dimensions. We discuss a possible interpretation of these corrections in terms of a gravitational theory violating the local Lorentz invariance in section~\ref{Hamiltonian}. Lastly, section~\ref{discussion} focuses on possible physical interpretation of our results and outlines the future developments.

Unless specified otherwise, we work in $D=4$ spacetime dimensions. We choose the mostly positive metric signature $(-,+,+,+)$.To keep track of quantum and gravitational effects, we maintain $\hbar$ and $G$ explicit, but we set $c=k_{\text{B}}=1$. Other conventions follow~\cite{MTW}.

\section{Einstein equations from thermodynamics}
\label{Einstein}

We briefly recall how the equilibrium conditions for locally constructed causal diamonds encode the traceless Einstein equations, before introducing higher order corrections. Although we are going to restrict ourselves to the vacuum case in the next section, here we take into account the presence of matter to show how it enters into the equations.

\subsection{Light-cone cut local causal diamonds}

We start by introducing the specific realisation of local, observer-dependent causal horizons we work with in this paper: the causal diamonds. In flat spacetime a causal diamond can be simply defined as the domain of dependence of a spacelike co-dimension~$1$ ball. In a generic curved spacetime only small causal diamonds can be meaningfully defined. More specifically, one requires that the geodesic radius of the co-dimension~$1$ ball forming the base of the causal diamond is much smaller than the local curvature length scale (i.e. the inverse of the square root of the largest eigenvalue of the Riemann tensor). Then, the spacetime curvature effects on the geometry of the causal diamonds can be treated as small corrections. However, even in this case, there exist several inequivalent generalisations of a causal diamond~\cite{Wang:2019}. While we could in principle consider any of them, the definition best suited for our purposes is the light-cone cut local causal diamond, as we explain in the following.

\begin{figure}[tbp]
    \centering
    \includegraphics[width=.45\textwidth,origin=c,trim={6.3cm 1.1cm 12.1cm 0.6cm},clip]{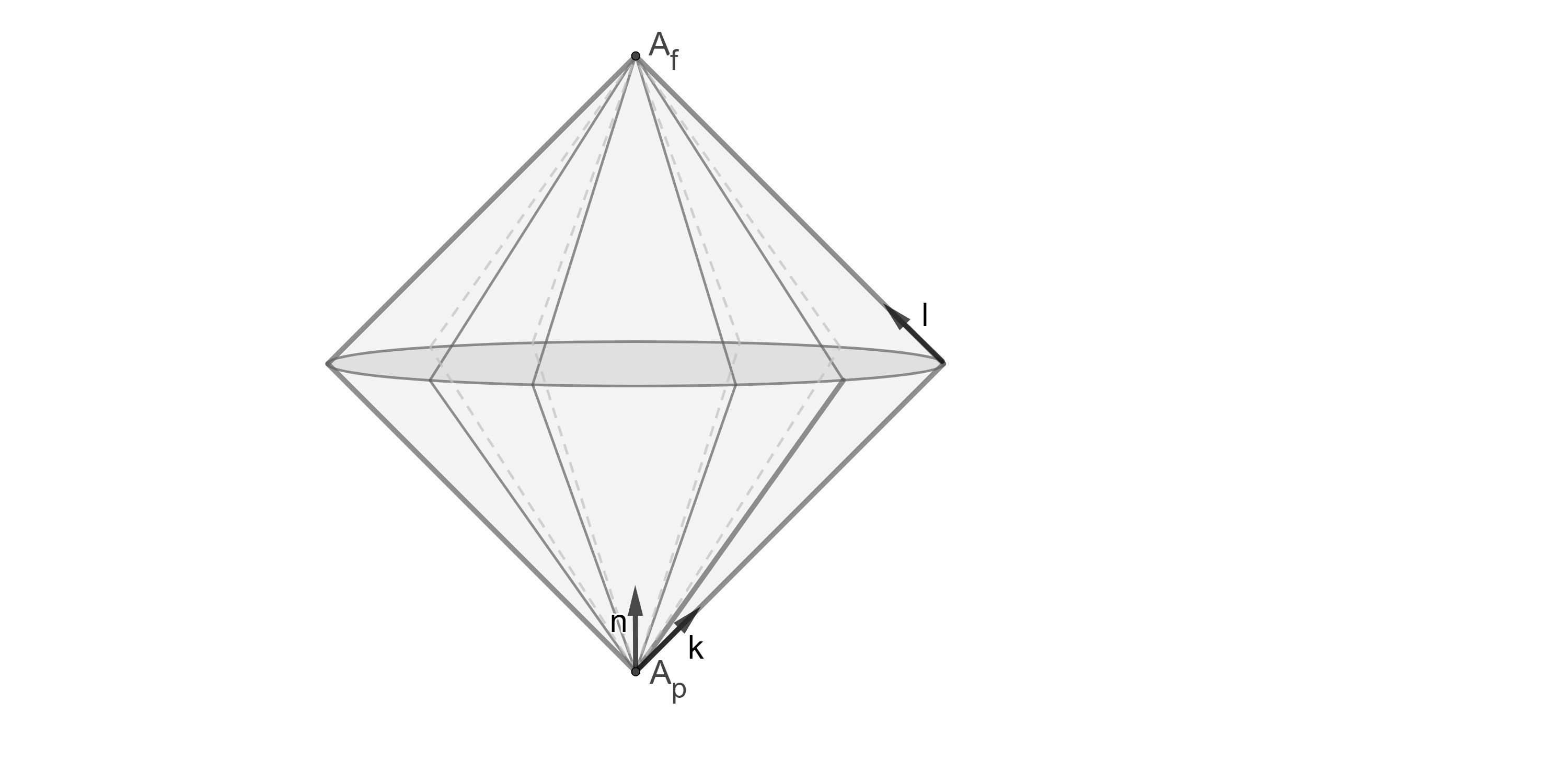}
    \caption{\label{diamond}{A sketch of a light-cone cut causal diamond starting in the past apex $A_{\text{f}}$. We suppress the $\theta$ coordinate. The unit, future-directed timelike vector $n^{\mu}$ defines the local direction of time. The past null boundary of the causal diamond is a congruence of null geodesics tangent to the future-pointing null vector $k^{\mu}$. Likewise, the future null boundary is a congruence tangent to the future-pointing null vector $l^{\mu}$. Several sample null geodesics forming the boundary are shown.}}
\end{figure}

The construction of a light-cone cut local causal diamond starts at an arbitrary regular spacetime point $A_{\text{p}}$, the past apex of the eventual causal diamond. At this point, we choose a unit timelike vector $n^{\mu}$ to represent a local direction of time. We then consider a class of future directed null vectors $k^{\mu}$ at $A_{\text{p}}$ normalised so that $n_{\mu}k^{\mu}=-1$, and define the past boundary of the light-cone cut local causal diamond as the congruence of null wordlines tangent to $k^{\mu}$. Any spacelike cross-section of this congruence is an approximate $2$-sphere (up to curvature-dependent corrections) and its interior an approximate $3$-dimensional spacelike ball. The base of the causal diamond corresponds to its spatial slice $\Sigma_0$ at the affine geodesic parameter length $l$. To construct the future (contracting) part of the light-cone cut local causal diamond, we introduce a future-directed null vector field $l^{\mu}$ defined on the boundary of $\Sigma_0$, so that $n_{\mu}l^{\mu}=-1$. Furthermore, denoting the projection of $k^{\mu}$ on the surface orthogonal to $n^{\mu}$ by $m^{\mu}$ (this vector is simply a unit, outward-pointing radial vector), we choose $l^{\mu}$ to have this projection equal to $-m^{\mu}$. Then, the contracting congruence of the null geodesics tangent to $l^{\mu}$ makes up the future boundary of the causal diamond. We show a sketch of the light-cone cut local causal diamond in figure~\ref{diamond}.

Since the size parameter $l$ of the light-cone cut local causal diamond is taken to be much smaller than the local curvature length scale, we can conveniently expand the metric inside the diamond using the Riemann normal coordinates, keeping track only of the lowest order terms. We obtain
\begin{equation}
\label{RNC}
g_{\mu\nu}(x)=\eta_{\mu\nu}-\frac{1}{3}R_{\mu\alpha\nu\beta}\left(A_{\text{p}}\right)x^{\alpha}x^{\beta}+O\left(x^3\right),
\end{equation}
where the Riemann tensor is evaluated at the past apex $A_{\text{p}}$ and $\eta_{\mu\nu}$ denotes the flat spacetime metric. A light-cone cut local causal diamond is constructed as an intersection of light cones, whose shape is invariant under conformal rescaling of the metric. Consequently, there exists an approximate conformal isometry (up to $O\left(l^3\right)$ curvature-dependent corrections) of the diamond, generated by a conformal Killing vector which reads
\begin{equation}
\label{conformal Killing}
\zeta^{\mu}=C\left[\left(l^2-t^2-r^2\right)n^{\mu}-2rtm^{\mu}\right],
\end{equation}
where $r$ is the radial geodesic distance, $t$ the affine time parameter measured along vector $n^{\mu}$, and $C$ stands for an arbitrary constant determining the normalisation of $\zeta^{\mu}$ with the dimensions of inverse length squared. The conformal Killing vector $\zeta^{\mu}$ is timelike inside the causal diamond and becomes null on its boundary. Thence, the boundary is a conformal Killing horizon.

The geometry of causal diamonds makes their thermodynamics particularly simple in several regards. First, a spatial (orthogonal to $n^{\mu}$) cross-section of the horizon of a causal diamond is closed, spherically symmetric and has a finite area. Therefore, it possesses a clearly defined interior region and, furthermore, there are no edges that might contribute to the horizon entropy. Second, the conformal Killing isometry of the causal diamonds allows for a very straightforward application of the covariant phase space methods~\cite{Bueno:2017,Jacobson:2019} (although we do not follow this route in the present work).

The light-cone cut local causal diamonds then offer the additional advantage of fully specifying the null boundary of the causal diamond (up to some residual freedom~\cite{Svesko:2019,Alonso:2024} we discuss in the following), while keeping the geometry of its spatial cross-sections somewhat ambiguous~\cite{Wang:2019}. Then, it becomes very simple to track the physical matter entropy flux across the boundary and the corresponding changes in its area. Since, as we review in the following, the equilibrium conditions imposed on these two processes encode the equations governing the gravitational dynamics, light-cone cut local causal diamonds are ideally suited for deriving them in this way.

\subsection{Gravitational dynamics from thermodynamics}

We have seen that the null boundary of a light-cone cut causal diamond represents a conformal Killing horizon. Then, it can be argued to possess entropy proportional to the area $\mathcal{A}$ of its spatial cross-section, i.e., $S=\eta\mathcal{A}$, where $\eta$ denotes the proportionality constant. This entropy can be interpreted as the von Neumann entropy sourced of the quantum entanglement between the two regions separated by the horizon~\cite{Sorkin:1986,Srednicki:1993,Solodukhin:2011}. However, more generally, it may also be understood as Shannon entropy accounting for the lack of information the observer inside the diamond has about its exterior~\cite{Bekenstein:1973}. Interested reader can find a more detailed discussion of this issue with further references here~\cite{Alonso:2024}. For the purposes of the present work, we do not care about the microscopic origin of this entropy, we simply take it as our assumption. The entropy, of course, changes as the area of the horizon expands. At the same time, matter flux crossing the horizon also carries entropy with it. It has been shown that the balance between both entropies encodes the gravitational dynamics~\cite{Svesko:2019,Alonso:2024}. In the following, we review the main points of this argument, first recalling the expressions for the changes of both relevant entropies.

To compute the matter entropy flux, we employ the Clausius equilibrium relation $\Delta S_{\text{C}}=\Delta Q/T_{\text{U}}$. Here, $T_{\text{U}}$ denotes the Unruh temperature measured by accelerating observers moving inside the causal diamond and $\Delta Q$ the heat flux perceived by the same observer. The total flux of this Clausius entropy over the past boundary of the causal diamond equals (for details on its definition and computation, see~\cite{Baccetti:2013ica})
\begin{equation}
\Delta S_{\text{C}}=\frac{2\pi}{\hbar}\frac{4\pi l^4}{9}T_{\mu\nu}\left(n^{\mu}n^{\nu}+\frac{1}{4}g^{\mu\nu}\right)+O\left(l^{6}\right), \label{Delta S_C}
\end{equation}
where $O\left(\epsilon^2l^D\right)$ accounts for the subleading corrections appearing due to the spacetime curvature and to the approximation of the energy-momentum tensor by its value at the diamond's past apex $A_{\text{f}}$.

We mentioned that the entropy of the horizon is proportional to its area. Its changes are then encoded in the expansion $\theta=\nabla_{\mu}k^{\mu}$ of the congruence of its geodesic generators, i.e.,
\begin{equation}
\label{area change}
\Delta S=\eta\Delta\mathcal{A}=\eta\int\theta\text{d}^3\Sigma,
\end{equation}
where, in our case, we integrate over the entire past boundary of the causal diamond. The expansion evolves according to the Raychaudhuri equation~\cite{Raychaudhuri:1955}
\begin{equation}
\label{Raychaudhuri}
\dot{\theta}=-\frac{1}{2}\theta^2-\sigma^2-R_{\mu\nu}k^{\mu}k^{\nu},
\end{equation}
where we introduced the simplified notation $\dot{\theta}=\text{d}\theta/\text{d}\lambda=k^{\mu}\nabla_{\mu}\theta$, with $\lambda$ being the affine parameter along the  geodesic null generators of the horizon. For computational convenience, we shift $\lambda$ by a constant, so that $\lambda=0$ at the base $\Sigma_0$ of the causal diamond. The second term on the right hand side of the Raychaudhuri equation~\eqref{Raychaudhuri} is given by the shear tensor \mbox{$\sigma^2=\sigma_{\mu\nu}\sigma^{\mu\nu}$}, which reads
\begin{equation}
\sigma_{\mu\nu}=h_{\mu}^{\;\:\lambda}h_{\nu}^{\;\:\rho}\nabla_{(\lambda}k_{\rho)}-\frac{1}{2}\nabla_{\rho}k^{\rho}h_{\mu\nu},
\end{equation}
where $h_{\mu\nu}$ denotes the induced metric on the null boundary (the shear is independent of its precise choice). The evolution of the shear tensor obeys the following equation~\cite{Raychaudhuri:1955}
\begin{equation}
\label{sigma}
\dot{\sigma}_{\mu\nu}=-\theta\sigma_{\mu\nu}-\sigma_{\mu\lambda}\sigma_{\nu}^{\;\:\lambda}+\frac{1}{2}\sigma^2h_{\mu\nu}-C_{\lambda\rho\sigma\tau}k^{\lambda}k^{\sigma}h_{\mu}^{\rho}h_{\nu}^{\tau}+\frac{1}{2}\left(h_{\mu\lambda}h_{\nu\rho}-\frac{1}{2}h_{\mu\nu}h_{\lambda\rho}\right)R^{\lambda\rho},
\end{equation}
where $C_{\lambda\rho\sigma\tau}$ is the Weyl curvature tensor.

To solve the coupled equations for the expansion~\eqref{Raychaudhuri} and the shear~\eqref{sigma}, we expand them in powers of $\lambda$. Since, on the past null horizon, $\lambda\in\left[-l,0\right]$, and we chose $l$ to be much smaller than the local curvature length scale, we can neglect the terms with high enough powers of $\lambda$, simplifying the calculations. We first have to determine the situation in flat spacetime. While the shear vanishes there, the horizon of a flat spacetime causal diamond expands at a rate
\begin{equation}
\theta_{\text{flat}}=\frac{2}{l+\lambda}.
\end{equation}
Therefore, in a curved spacetime, we can take the following ansatze for the expansion and the shear
\begin{align}
\theta&=\theta_{\text{flat}}+\theta_{(0)}+\lambda\theta_{(1)}+O\left(\lambda^2\right), \label{Raych} \\
\sigma_{\mu\nu}&=\sigma_{(0)\mu\nu}+\lambda\sigma_{(1)\mu\nu}+O\left(\lambda^2\right). \label{shear}
\end{align}
Since, by definition, the diamond expands to the past of the base $\Sigma_0$ and contracts to the future of it, the smooth evolution of the expansion requires $\theta_{(0)}=0$. It has been shown that non-zero $\sigma_{0,\mu\nu}$ leads to non-equilibrium entropy production~\cite{Chirco:2010,Chirco:2010b} analogous to the tidal heating of black holes~\cite{Hawking:1972}. This non-equilibrium entropy production can be straightforwardly dealt with, treating it in the same way as the $\theta_{\text{flat}}$ contribution we discuss in the following. Nevertheless, for simplicity, we set here $\sigma_{0\mu\nu}=0$. Without this technical condition, everything proceeds in the same way, but the higher order calculations in the next section become needlessly messy. Upon fixing $\sigma_{0\mu\nu}=0$, it is easy to see that the shear contributions to the expansion are not relevant and we obtain the following solution
\begin{align}
\theta&=\theta_{\text{flat}}-\lambda R_{\mu\nu}k^{\mu}k^{\nu}+O\left(\lambda^2\right). \label{expansion}
\end{align}

Expansion~\eqref{expansion} plugged into equation~\eqref{area change} directly determines the change of entropy along the past boundary of the causal diamond. We obtain
\begin{equation}
\Delta S=\eta\int\theta_{\text{flat}}\text{d}^3\Sigma-\eta\int\lambda R_{\mu\nu}k^{\mu}k^{\nu}\text{d}^3\Sigma+O\left(l^5\right).
\end{equation}
The first term is dependent on the flat spacetime expansion of the causal diamond. Therefore, the corresponding change in entropy remains non-zero even in the absence of any matter Clausius entropy flux compensating it. Such a term then cannot describe a reversible thermodynamic process. It has been proposed that the flat spacetime expansion of the horizon corresponds to an irreversible process, akin to a free expansion of gas released from a container~\cite{Svesko:2018}. Therefore, we discard the entire contribution proportional to $\theta_{\text{flat}}$ as non-equilibrium, irreversible entropy production, keeping only the reversible part
\begin{equation}
\Delta S_{\text{rev}}=-\eta\int\lambda R_{\mu\nu}k^{\mu}k^{\nu}\text{d}^3\Sigma+O\left(l^5\right).
\end{equation}
We can carry out the integration, using $k^{\mu}=n^{\mu}+m^{\mu}$ and the well known results for integrating unit radial vectors over a $2$-sphere
\begin{align}
\int m^{\mu}\text{d}\Omega_{2}&=0, \\
\int m^{\mu}m^{\nu}\text{d}\Omega_{2}&=\frac{4\pi}{3}\gamma^{\mu\nu}.
\end{align}
where $\text{d}\Omega_{2}=\sin\theta\text{d}\theta\text{d}\phi$ is the angular integration element on the $2$-sphere and we introduced the spatial metric $\gamma^{\mu\nu}=g^{\mu\nu}+n^{\mu}n^{\nu}$. The result reads
\begin{equation}
\Delta S_{\text{rev}}=\frac{4\pi l^4}{9}R_{\mu\nu}\left(n^{\mu}n^{\nu}+\frac{1}{4}g^{\mu\nu}\right)+O\left(l^{5}\right). \label{Delta S rev}
\end{equation}

The thermodynamic equilibrium condition for the light-cone cut local causal diamond then implies that the reversible change of the horizon entropy $\Delta S_{\text{rev}}$ is exactly compensated by the matter Clausius entropy flux $\Delta S_{\text{C}}$, i.e., $\Delta S_{\text{rev}}+\Delta S_{\text{C}}=0$. This condition yields the following requirement
\begin{equation}
\label{precursor}
\left(R_{\mu\nu}-\frac{2\pi}{\hbar\eta}T_{\mu\nu}\right)\left(n^{\mu}n^{\nu}+\frac{1}{4}g^{\mu\nu}\right)=0.
\end{equation}
Any unit, future-pointing timelike vector $n^{\mu}$ defined in $A_{\text{p}}$ can be used as the local direction of time in a construction of a light-cone cut local causal diamond. For any such diamond, we can derive equation~\eqref{precursor}, just with a different timelike vector. Therefore, equation~\eqref{precursor} holds for an arbitrary unit, future-pointing timelike vector $n^{\mu}$. Then, it implies
\begin{equation}
\label{precursor 2}
R_{\mu\nu}-\frac{1}{4}Rg_{\mu\nu}=\frac{2\pi}{\hbar\eta}\left(T_{\mu\nu}-\frac{1}{4}Tg^{\mu\nu}\right).
\end{equation}
We provide a short proof of a this claim (actually of its more general version) in~\ref{n-dependence}.

Equations~\eqref{precursor 2} closely resemble the traceless Einstein equations. Taking their Newtonian limit allows us to identify the Newton gravitational constant $G$. Then, $G$ is defined in terms of the Planck constant $\hbar$ and the entropy proportionality constant $\eta$, i.e., $G=1/\left(4\hbar\eta\right)$. Then, we indeed recover the traceless Einstein equations\footnote{Let us note that there exists non-metric, first order actions leading to dynamics equivalent to that given by equation~\eqref{traceless}, notably the Plebański-like action~\cite{Montesinos:2023,Montesinos:2025} and the geometric trinity formulation of unimodular gravity~\cite{Nakayama:2023}. Whether such first order dynamics can be recovered from thermodynamic equilibrium conditions remains an open question. Nevertheless, thermodynamics of first order theories is well established~\cite{Koivisto:2022} and it should be possible to use it to recover the corresponding equations of motion.}
\begin{equation}
\label{traceless}
R_{\mu\nu}-\frac{1}{4}Rg_{\mu\nu}=8\pi G\left(T_{\mu\nu}-\frac{1}{4}Tg^{\mu\nu}\right).
\end{equation}
By virtue of the strong equivalence principle, these equations are valid throughout the spacetime~\cite{Jacobson:1995ab,Chirco:2010,Alonso:2024}. As an aside, the relation between $G$, $\hbar$ and $\eta$ we derived implies $\eta=1/\left(4l_{\text{P}}^2\right)$, with $l_{\text{P}}=\sqrt{\hbar G}$ being the Planck length. Thence, entropy of a local causal horizon has the same form as Bekenstein entropy of a black hole, $S=\mathcal{A}/\left(4l_{\text{P}}^2\right)$.

Let us conclude with two short remarks regarding the interpretation of equations~\eqref{traceless}. First, they are traceless, with the cosmological constant appearing only as an on-shell integration constant. Moreover, the geometry of the light-cone cut local causal diamonds is insensitive to the overall conformal factor of the metric. We have previously shown that these features imply that the gravitational dynamics we recover from thermodynamics (without any additional assumptions) does not correspond to general relativity, but rather to Weyl transverse (unimodular) gravity~\cite{Alonso:2020,Alonso:2021,Alonso:2024}. This theory has the same classical solutions as general relativity~\cite{Finkelstein:2001,Alvarez:2006,Carballo:2022}. Moreover, the quantum dynamics of both theories have so far been found equivalent~\cite{Carballo:2022,Kugo:2021,Kugo:2022,Garcia:2023b}. The important difference is that in Weyl transverse gravity the cosmological constant is not sourced by the vacuum fluctuations, which makes its value radiatively stable~\cite{Finkelstein:2001,Unruh:1989,Ellis:2011,Ellis:2014,Barcelo:2014,Carballo:2015,Barcelo:2018}. Note that vacuum fluctuations not proportional to the metric tensors do gravitate, allowing for quantum gravitational corrections to the equations of motion of the kind described in the next section.

Therefore, the corrections to gravitational dynamics we discuss in the following should also be seen as modifying Weyl transverse gravity rather than general relativity.

Second, the fact that thermodynamics of local causal horizons recovers the equations governing gravitational dynamics is often interpreted as suggesting that gravity is emergent. In other words, the Einstein equations would be essentially equations of state describing the behaviour of some unknown quantum degrees of freedom of the spacetime in the classical limit~\cite{Jacobson:1995ab,Padmanabhan:2010,Chirco:2010,Verlinde:2011}. While we find this viewpoint worthwhile and thought provoking, we believe it is simply unnecessary to embrace it. We have shown that thermodynamics leads to novel insights into gravitational dynamics, even without requiring that the latter is emergent~\cite{Alonso:2020,Alonso:2021,Alonso:2024,Alonso:2023}. All we need to assume is that the local equilibrium conditions encode \textit{all} the information necessary to reconstruct the gravitational dynamics. We maintain this viewpoint in the rest of the paper.

\section{Higher order corrections in vacuum: the Bel-Robinson tensor}	
\label{Weyl corrections}

In the previous section, we have reviewed the derivation of the traceless Einstein equations from thermodynamics. We neglected the subleading contributions to the curved spacetime changes of the area of the spatial cross-section of the horizon, and, hence, to its entropy. It is natural to ask whether these terms encode some corrections to Einstein equations. In some special cases, we have previously shown that this is indeed the case. However, such corrections only occur if we also take into account the subleading quantum contribution to the horizon entropy proportional to the logarithm of its area~\cite{Alonso:2023,Alonso:2020b}. Specifically, we have shown that the derivation generically leads to corrections to the traceless Einstein equations quadratic in the Ricci tensor~\cite{Alonso:2020b,Alonso:2023b}. We have also derived the linearised modified dynamics, finding a result equivalent to a special case of linearised quadratic gravity~\cite{Alonso:2023}. Due to various technical and conceptual challenges, we are for the time unable to derive the modified equations in the fully general case. Here, we instead study another simplified setting, the vacuum case. On the one hand, we use this setting to develop tools that should allow us to derive the completely general equations in a subsequent work. On the other hand, any possible modifications to vacuum gravitational dynamics are of great interest by themselves, being the only kind of corrections relevant to the standard black hole solutions, in particular, to the Kerr metric (see~\cite{Hennigar:2016,Bueno:2024} for examples of such corrections). They are also important for the behaviour of the spacetime in the vicinity of generic spacelike singularities, as described by the Belinskii-Khalatnikov-Lifshitz conjecture~\cite{Belinskii:1971}. Finally, such corrections generically modify the propagation of gravitational waves, which might make them experimentally falsifiable by the current or future gravitational waves detectors~\cite{Zumalacarregui:2017}. To study possible vacuum corrections to gravitational dynamics implied by the thermodynamic derivation, we can follow the same basic strategy we employed to derive the traceless Einstein equations. However, we need to introduce some conceptual changes, both on the level of the initial assumptions and in the process of the derivation. Therefore, we start by reviewing the requirements we impose.

In the semiclassical setup, we have relied on the Einstein equivalence principle to construct the local Minkowski vacuum necessary for invoking the Unruh effect. However, the status of any formulation of the equivalence principle in a regime in which quantum gravitational corrections become relevant remains an open issue. Therefore, we now instead impose two more modest requirements, which do not rely on the general formulation of the equivalence principle but still allow for a thermodynamic description of local causal horizons. Our first requirement then concerns an observer equipped with a particle detector and moving at an approximately constant acceleration. Given enough time for the detector to thermalise, we require that it registers local Minkowski vacuum as a thermal bath of particles at the Unruh temperature $T_{\text{U}}=a/\left(2\pi\right)$. While the fate of the Unruh effect in quantum gravity remains an open questions, there exist indications that it remains viable~\cite{Carballo:2019}.

As an aside, in the context of the generalised uncertainty principle phenomenology~\cite{Garay:1994}, it has been proposed that the Unruh temperature might develop corrections due to the interplay of quantum and gravitational physics~\cite{Scardigli:2018}. It remains an open question whether such corrections can really occur. In any case, in a previous work, we have shown that their presence in any case cannot affect the gravitational dynamics~\cite{Alonso:2020b}. Therefore, we can safely proceed with the standard Unruh temperature, as any potential subleading corrections to it are irrelevant.

The second requirement we impose is a modified entropy which includes the leading order quantum corrections including corrections from quantum contributions\footnote{Note here, that it is possible to explore more radical modifications of entropy, including non-extensive proposals, and study the possible modified gravitational dynamics they encode. While this option constitutes a natural extension of our research programme, we do not pursue it here.}. In particular, entropy of a  spherically symmetric local causal horizon in four spacetime dimensions has a generic form
\begin{equation}
\label{log entropy}
S=\frac{\mathcal{A}}{4l_{\text{P}}^2}+\mathcal{C}\ln\frac{\mathcal{A}}{\mathcal{A}_0}+O\left(\frac{\mathcal{A}_0}{\mathcal{A}}\right),
\end{equation}
where $\mathcal{A}_0$ is a constant with the dimensions of area and $\mathcal{C}$ a real number. Both $\mathcal{A}_0$ and $\mathcal{C}$ are characteristic for the particular approach to computing entropy, but independent of the horizon for which it is computed. The first two terms in equation~\eqref{log entropy} remain in the limit $l_{\text{P}}\to0$ (and the leading one diverges), whereas the $O\left(\mathcal{A}_0/\mathcal{A}\right)$ contributions vanish. Then, restricting our attention to the area term and the logarithm is equivalent to keeping the only two contributions to entropy relevant for large horizons, such that $\mathcal{A}\gg l_{\text{P}}^2$.

Ideally, parameter $\mathcal{C}$ could be measured in some experimental set up to detect quantum gravity corrections. However, like other quantum gravitational corrections, the value of C is beyond the current experimental limits. Nevertheless, there already exist proposals to establish bounds on $\mathcal{C}$~\cite{Alonso:2021b} and quantify the changes this parameter implies, e.g. for the thermal spectrum of particles produced by the Hawking effect~\cite{Alonso:2018}.

Equation~\eqref{log entropy} arises independently in a number of different contexts. It appears, e.g. in loop quantum gravity~\cite{Kaul:2000,Meissner:2004}, string theory~\cite{Banerjee:2011,Sen:2013,Karan:2023}, AdS/CFT correspondence~\cite{Faulkner:2013}, entanglement entropy calculations~\cite{Solodukhin:2011,Solodukhin:2010}, phenomenological approaches based on the existence of a minimal length~\cite{Adler:2001,Medved:2004}, in an analysis of statistical fluctuations~\cite{Medved:2005,Das:2002,Gour:2003} and from quantisation of the horizon area~\cite{Hod:2004,Davidson:2019}. In other words, the logarithmic correction to entropy associated with causal horizons is present in most of the approaches to quantum gravity. The specific approach is characterised just by the sign and value of the parameter $\mathcal{C}$. Thence, the modifications of gravitational dynamics implied by the logarithmic term are in principle relevant for any approach to quantum gravity. They are only parametrised by a single unknown number, $\mathcal{C}$, which is of the order of unity for most approaches.

Upon introducing the initial assumptions, we move on to the derivation of the equations governing gravitational dynamics. Since we work in vacuum, there is no matter Clausius entropy flux present. Therefore, our task is simply to compute the total change in the area of the horizon. It will consist of two components, one, $\Delta A_{\text{irr}}$ proportional to $\theta_{\text{flat}}$ and, thence, corresponding to an irreversible thermodynamic process, and the remaining reversible part $\Delta A_{\text{rev}}$. If the horizon is in equilibrium, the reversible change in its entropy must vanish. As we saw in the previous section, this condition then encodes the equations governing gravitational dynamics, in this case including the potential vacuum corrections.

We now need to solve the Raychaudhuri equation~\eqref{Raychaudhuri} and equation~\eqref{shear} for the evolution of the shear tensor to the order $O\left(\lambda^3\right)$ (recall that the affine null geodesic parameter $\lambda$ lies in the interval from $-l$ to $0$ on the past boundary), which becomes relevant once we take into account the logarithmic corrections to entropy. We obtain (keeping in place the condition $\theta_0=\sigma_{0\mu\nu}=0$)
\begin{align}
\nonumber \theta\left(x\right)=&\theta_{\text{flat}}\bigg[1+\frac{\lambda^2}{1+\lambda\theta_{\text{flat}}}R_{\mu\nu}k^{\mu}k^{\nu}+\frac{1}{6}\frac{2+\lambda\theta_{\text{flat}}}{\left(1+\lambda\theta_{\text{flat}}\right)^2}\lambda^4\left(R_{\mu\nu}k^{\mu}k^{\nu}\right)^2\bigg] \\
&-\lambda R_{\mu\nu}k^{\mu}k^{\nu}-\frac{1}{6}\lambda^3\left(R_{\mu\nu}k^{\mu}k^{\nu}\right)^2-\frac{1}{3}\lambda\sigma^2+O\left(\lambda^4\right), \\
\nonumber \sigma_{\mu\nu}\left(x\right)=&\lambda\left[C_{\lambda\rho\sigma\tau}k^{\lambda}k^{\sigma}h_{\mu}^{\rho}h_{\nu}^{\tau}-\frac{1}{2}\left(h_{\mu\lambda}h_{\nu\rho}-\frac{1}{2}h_{\mu\nu}h_{\lambda\rho}\right)R^{\lambda\rho}\right]\left(-1+\frac{\lambda\theta_{\text{flat}}}{1+\lambda\theta_{\text{flat}}}\right) \\
&+O\left(\lambda^2\right),
\end{align}
where we evaluate all the quantities at an arbitrary point at the past horizon of the causal diamond, given by coordinates $x^{\mu}=\left(l+\lambda\right)k^{\mu}$. While we can compute the shear tensor up to $O\left(\lambda^3\right)$ as well, the subleading terms only affect the evolution of the expansion at the order $O\left(\lambda^4\right)$, which we neglect.

Not all the terms contributing to $\theta$ are relevant in vacuum, if we are interested in analysing perturbative corrections to gravitational dynamics. From the form of entropy with the logarithmic term~\eqref{log entropy}, it is easy to realise that the corrections are going to be suppressed by $l_{\text{P}}^2$ and that we neglect any $O\left(l_{\text{P}}^4\right)$ corrections. To the order $O\left(l_{\text{P}}^0\right)$ the traceless Einstein equations imply that the Ricci tensor is given by an integration constant, $R_{\mu\nu}=\Lambda g_{\mu\nu}$ and the combination $R_{\mu\nu}k^{\mu}k^{\nu}$ therefore vanishes, since $k^{\mu}$ is a null vector. Consequently, the value of $R_{\mu\nu}k^{\mu}k^{\nu}$ must be $O\left(l_{\text{P}}^2\right)$, as it can only be sourced by the vacuum corrections to the traceless Einstein equations. This observation allows us to neglect the terms quadratic in the Ricci tensor as it must be $O\left(l_{\text{P}}^4\right)$. By the same reasoning, we can also drop the terms containing a product of the Ricci and the Weyl tensors.

The irreversible change in the area is given by the integral of the part of $\theta$ proportional to the flat spacetime expansion $\theta_{\text{flat}}$, i.e. (we drop the aforementioned irrelevant contributions),
\begin{equation}
\Delta\mathcal{A}_{\text{irr}}=\int\theta_{\text{flat}}\left[1+\frac{\lambda^2}{1+\lambda\theta_{\text{flat}}}R_{\mu\nu}k^{\mu}k^{\nu}+\frac{1}{3}\frac{\lambda^5\theta_{\text{flat}}}{\left(1+\lambda\theta_{\text{flat}}\right)^2}\left(C_{\lambda\rho\sigma\tau}k^{\lambda}k^{\sigma}h_{\mu}^{\rho}h_{\nu}^{\tau}\right)^2\right]\text{d}^3\Sigma.
\end{equation}
The remainder of the expansion then corresponds to the reversible change in the area, which reads
\begin{equation}
\Delta\mathcal{A}_{\text{rev}}=-\int\left[\lambda R_{\mu\nu}k^{\mu}k^{\nu}+\frac{1}{3}\lambda^3\left(C_{\lambda\rho\sigma\tau}k^{\lambda}k^{\sigma}h_{\mu}^{\rho}h_{\nu}^{\tau}\right)^2\right]\text{d}^3\Sigma.
\end{equation}
To carry out the integrations, we need to expand the tensors around their value at the past apex $A_{\text{f}}$. The derivative terms in such expansion of the term quadratic in the Weyl tensor contribute to the integrand at the order $O\left(\lambda^4\right)$ and can be neglected. Nevertheless, the contributions of the derivatives of the Ricci tensor are in principle relevant. However, any such term would eventually end up in the modified Einstein equations as a source term for the Ricci tensor suppressed by a factor $l_{\text{P}}^2$. Since the value of the Ricci tensor in vacuum is already of the order $O\left(l_{\text{P}}^2\right)$, any terms with the derivatives of the Ricci tensor contribute to the equations governing the gravitational dynamics only at the order $O\left(l_{\text{P}}^4\right)$. Therefore, in vacuum, we can safely neglect any such terms and approximate the curvature tensors in the integrands just by their value at $A_{\text{f}}$.

Plugging $\Delta\mathcal{A}_{\text{irr}}$ and $\Delta\mathcal{A}_{\text{rev}}$ into equation~\eqref{log entropy} for the entropy with a logarithmic correction, allows us to find the expression for the reversible change of entropy
\begin{align}
\nonumber \Delta S_{\text{rev}}=&\Delta S_{\text{total}}-\Delta S_{\text{irr}} \\
\nonumber  =& \frac{\Delta\mathcal{A}_{\text{irr}}+\Delta\mathcal{A}_{\text{rev}}}{4l_{\text{P}}^2}+\mathcal{C}\ln\frac{\Delta\mathcal{A}_{\text{irr}}+\Delta\mathcal{A}_{\text{rev}}}{\mathcal{A}_0}-\frac{\Delta\mathcal{A}_{\text{irr}}}{4l_{\text{P}}^2}-\mathcal{C}\ln\frac{\Delta\mathcal{A}_{\text{irr}}}{\mathcal{A}_0} \\
=&\frac{\Delta\mathcal{A}_{\text{rev}}}{4l_{\text{P}}^2}+\mathcal{C}\ln\left(1+\frac{\Delta\mathcal{A}_{\text{rev}}}{\Delta\mathcal{A}_{\text{irr}}}\right).
\end{align}
The logarithmic term can be expanded, yielding
\begin{equation}
\Delta S_{\text{rev}}=\frac{\Delta\mathcal{A}_{\text{rev}}}{4l_{\text{P}}^2}+\mathcal{C}\frac{\Delta\mathcal{A}_{\text{rev}}}{\Delta\mathcal{A}_{\text{irr}}}+O\left(l^6\right),
\end{equation}
using that the quadratic term contributes only at the order $O\left(l^6\right)$ in vacuum (but not in the presence of matter~\cite{Alonso:2020b}).

Let us now explicitly compute $\Delta\mathcal{A}_{\text{irr}}$ and $\Delta\mathcal{A}_{\text{rev}}$. For $\Delta\mathcal{A}_{\text{irr}}$ it is easy to realise that any terms proportional to the Ricci tensor lead to contributions to $\Delta S_{\text{rev}}$ quadratic in the Ricci tensor which affect the gravitational dynamics only at the order $O\left(l_{\text{P}}\right)^4$ and can be safely discarded. It is easy to check that the leading order correction to the volume element on the horizon in the Riemann normal coordinates also goes with the Ricci tensor. Therefore, the only relevant contribution to $\Delta\mathcal{A}_{\text{irr}}$ is the flat spacetime one and we obtain
\begin{equation}
\Delta\mathcal{A}_{\text{irr}}=4\pi l^2.
\end{equation}
For $\Delta\mathcal{A}_{\text{rev}}$ the Ricci tensor correction to the integration elements also leads to terms quadratic in the Ricci tensor. Therefore, we can integrate with the flat spacetime volume element
\begin{equation}
\Delta\mathcal{A}_{\text{rev}}=-\int_{-l}^{0}\text{d}\lambda\int\text{d}\Omega_{2}\left(l+\lambda\right)^2\left[\lambda R_{\mu\nu}k^{\mu}k^{\nu}+\frac{1}{3}\lambda^3\left(C_{\lambda\rho\sigma\tau}k^{\lambda}k^{\sigma}h_{\mu}^{\rho}h_{\nu}^{\tau}\right)^2\right].
\end{equation}
We may decompose the vector field $k^{\mu}$ into $k^{\mu}=n^{\mu}+m^{\mu}$, with $n^{\mu}$ being the local direction of time and $m^{\mu}$ the unit radial spacelike vector. For the angular integral of the radial vectors $m^{\mu}$ over a sphere, we have~\cite{Jacobson:2017}
\begin{align}
\int m^{\mu}\text{d}\Omega_{2}=&0, \\
\int m^{\mu}m^{\nu}\text{d}\Omega_{2}=&\frac{4\pi}{3}\gamma^{\mu\nu}, \\
\int m^{\mu}m^{\nu}m^{\lambda}\text{d}\Omega_{2}=&0, \\
\int m^{\mu}m^{\nu}m^{\lambda}m^{\rho}\text{d}\Omega_{2}=&\frac{4\pi}{15}\left(\gamma^{\mu\nu}\gamma^{\lambda\rho}+\gamma^{\mu\lambda}\gamma^{\nu\rho}+\gamma^{\mu\rho}\gamma^{\nu\lambda}\right).
\end{align}
The integration in the null parameter $\lambda$ is straightforward. In the end, we arrive at the following expression for $\Delta\mathcal{A}_{\text{rev}}$
\begin{equation}
\Delta\mathcal{A}_{\text{rev}}=-\frac{4\pi l^4}{9}\left[\left(R_{\mu\nu}-\frac{1}{4}Rg_{\mu\nu}\right)n^{\mu}n^{\nu}+\frac{2}{25}l^2T_{\mu\nu\lambda\rho}n^{\mu}n^{\nu}n^{\lambda}n^{\rho}\right],
\end{equation}
where
\begin{align}
\nonumber T_{\mu\nu\lambda\rho}=&C_{\mu\sigma\lambda\tau}C_{\nu\;\:\rho}^{\;\:\sigma\;\:\tau}+C_{\mu\sigma\rho\tau}C_{\nu\;\:\lambda}^{\;\:\sigma\;\:\tau}-\frac{1}{2}g_{\mu\nu}C_{\lambda\sigma\alpha\beta}C_{\rho}^{\;\:\sigma\alpha\beta}-\frac{1}{2}g_{\lambda\rho}C_{\mu\sigma\alpha\beta}C_{\nu}^{\;\:\sigma\alpha\beta} \\
&+\frac{1}{8}g_{\mu\nu}g_{\lambda\rho}C_{\sigma\alpha\beta\gamma}C^{\sigma\alpha\beta\gamma}, \label{Bel-Robinson}
\end{align}
denotes the Bel-Robinson tensor~\cite{Bel:1958,Senovilla:2000}.

The reversible change of entropy then equals
\begin{align}
\nonumber \Delta S_{\text{rev}}=&-\frac{4\pi l^4}{36l_{\text{P}}^2}\left(R_{\mu\nu}-\frac{1}{4}Rg_{\mu\nu}\right)n^{\mu}n^{\nu}-\mathcal{C}\frac{l^2}{9}\bigg[\left(R_{\mu\nu}-\frac{1}{4}Rg_{\mu\nu}\right)n^{\mu}n^{\nu} \\
&+\frac{2}{25}l^2T_{\mu\nu\lambda\rho}n^{\mu}n^{\nu}n^{\lambda}n^{\rho}\bigg]+O\left(l^6\right). \label{dS rev}
\end{align}
Thus, the local equilibrium condition reads $\Delta S_{\text{rev}}=0$, i.e.,
\begin{equation}
\left(R_{\mu\nu}-\frac{1}{4}Rg_{\mu\nu}\right)n^{\mu}n^{\nu}+\frac{\mathcal{C}l_{\text{P}}^2}{\pi l^2}\left(R_{\mu\nu}-\frac{1}{4}Rg_{\mu\nu}\right)n^{\mu}n^{\nu} +\frac{2\mathcal{C}l_{\text{P}}^2}{25\pi}T_{\mu\nu\lambda\rho}n^{\mu}n^{\nu}n^{\lambda}n^{\rho}=0,
\end{equation}
where we discarded the $O\left(l^6\right)$ terms\footnote{Justifying the removal of these terms is in fact rather subtle. To treat the spacetime as a differentiable manifold, we must keep the size parameter of the causal diamond much larger than the Planck scale. Then, $O\left(l^6\right)$ contribution can actually be larger than the $O\left(l^4l_{\text{P}}^2\right)$ ones. However, the size parameter $l$ is only restricted to be much larger than the Planck length and at the same time much smaller than the local curvature length scale. Within this range, the value of $l$ is completely arbitrary. Therefore, at any regular spacetime point, we can construct a sequence of causal diamonds with different size parameters. For any such diamond, we can derive a version of equation~\eqref{dS rev}, differing only in the value of $l$. Consequently, equation~\eqref{dS rev} is only satisfied if the terms proportional to different powers of $l$ vanish separately. This observation allows us to focus on the leading order $O\left(l^4\right)$ part of this equation and disregard all the higher order contributions. For the details of this procedure see a previous work of the authors~\cite{Alonso:2023}.}. Since the Ricci tensor is $O\left(l_{\text{P}}^2\right)$, the second term only contributes at $O\left(l_{\text{P}}^4\right)$ and we can also discard it, leaving the following equation
\begin{equation}
\label{constraint}
\left(R_{\mu\nu}-\frac{1}{4}Rg_{\mu\nu}\right)n^{\mu}n^{\nu}=-\frac{2\mathcal{C}l_{\text{P}}^2}{25\pi}T_{\mu\nu\lambda\rho}n^{\mu}n^{\nu}n^{\lambda}n^{\rho}.
\end{equation}
To obtain this result, we only imposed the form of entropy with the logarithmic corrections~\eqref{log entropy}, the survival of the Unruh effect in the regime we discuss, and locality of the equations governing the gravitational dynamics. However, the physical interpretation of equation~\eqref{constraint} depends on additional assumptions, which we discuss in the next section.

\section{Interpretation of the Bel-Robinson tensor contribution}
\label{Hamiltonian}

We start by briefly reviewing the properties of the Bel-Robinson tensor $T_{\mu\nu\lambda\rho}$, which appears as the source term in equation~\eqref{constraint}. The Bel-Robinson tensor in $4$ spacetime dimensions (the only case we consider) is given by equation~\eqref{Bel-Robinson}. It is symmetric in all of its indices and traceless (these properties are specific to $4$ dimensions). In any spacetime dimension, it obeys an analogue of the dominant energy condition, i.e., for any four timelike, future-pointing vector $v_{i}^{\mu}$, $i\in\left[1,4\right]$, it holds $T_{\mu\nu\lambda\rho}v_1^{\mu}v_2^{\nu}v_3^{\lambda}v_4^{\rho}\ge0$ and the vector $-T_{\mu\nu\lambda\rho}v_1^{\nu}v_1^{\lambda}v_1^{\rho}$ is future-pointing. The Bel-Robinson tensor is non-degenerate, in the sense that it vanishes if and only if the Weyl tensor is equal to zero. Moreover, its divergence is identically zero in the Einstein spacetimes (i.e., the spacetimes with constant Ricci curvature). The Bel-Robinson tensor is in fact the unique tensor quadratic in the Weyl tensor with these properties~\cite{Senovilla:2000,Senovilla:1997,Senovilla:1999}.

The contraction of the Bel-Robinson tensor with a unit, future pointing, timelike vector in all indices (such as the term $T_{\mu\nu\lambda\rho}n^{\mu}n^{\nu}n^{\lambda}n^{\rho}$ we obtained in the previous section) is known as the Bel-Robinson super-energy density. This object appears in various different settings in the role corresponding to a quasilocal measure of the gravitational energy~\cite{Senovilla:2000,Szabados:2009}. In particular, the Bel-Robinson super-energy density determines the leading order vacuum deformation of the spatial volume of a small geodesic ball~\cite{Horowitz:1982,Bergqvist:1994}. In the presence of matter fields, this role is played by the energy density $T_{\mu\nu}n^{\mu}n^{\nu}$. Thence, it has been proposed that the Bel-Robinson super-energy density corresponds to the quasilocal gravitational energy density per unit area~\cite{Senovilla:2000,Garecki:1977} (for dimensional reasons).

A suggestive way to write the Bel-Robinson super-energy density comes from the so-called electric-magnetic decomposition of the Weyl tensor, which is named for being analogous to the decomposition of the Faraday tensor of the electromagnetic field~\cite{Senovilla:2001}. In $4$ dimensions, the ``electric'' part of the Weyl tensor reads
\begin{equation}
E_{\mu\nu}=C_{\lambda\alpha\rho\beta}u^{\lambda}u^{\rho}\gamma^{\alpha}_{\mu}\gamma^{\beta}_{\nu},
\end{equation}
and the magnetic part is
\begin{equation}
B_{\mu\nu}=\frac{1}{2}n^{\lambda}\varepsilon_{\lambda\mu\alpha\beta}C_{\rho\sigma}^{\;\:\;\:\alpha\beta}n^{\rho}\gamma^{\sigma}_{\nu},
\end{equation}
with $\varepsilon_{\lambda\mu\alpha\beta}$ being the Levi-Civita tensor. Written as a function of these tensors, the Bel-Robinson super-energy density becomes
\begin{equation}
\label{EM Bel}
T_{\mu\nu\lambda\rho}n^{\mu}n^{\nu}n^{\lambda}n^{\rho}=E_{\mu\nu}E^{\mu\nu}+B_{\mu\nu}B^{\mu\nu}=E^2+B^2.
\end{equation}
This expression is similar to the energy density of the electromagnetic field, although the analogy is of course purely formal.

In our case, the Bel-Robinson super-energy density multiplied by a Planck scale constant with the dimensions of area $-2\mathcal{C}l_{\text{P}}^2/\left(25\pi\right)$ appears as a source for the equations governing the gravitational dynamics. Therefore, it is tempting to interpret our result as the modified Einstein equations in which the Ricci curvature is sourced by the quasilocal gravitational energy density integrated over some Planck-scale unit of area. However, as we show in the following, embracing this interpretation requires that we surrender the local Lorentz invariance of physics at the Planck scale.

\subsection{Locally Lorentz invariant interpretation}

Let us first take the assumption of standard physics that the local Lorentz invariance is not violated by the corrections we introduce. This assumption is very natural, since the entire framework of the thermodynamics spacetime implicitly relies on it, especially in introducing the local causal horizons and the Unruh effect. It follows that no preferred direction of time can exist and the unit, timelike, future-pointing vector field $n^{\mu}$ present in equation~\eqref{constraint} is completely arbitrary (we discuss this arbitrariness in more detail in section~\ref{Einstein}). We can easily rewrite equation~\eqref{constraint} as a rank $4$ tensor contracted in all its indices with $n^{\mu}$ being equal to zero, i.e.,
\begin{equation}
\left[\left(R_{\mu\nu}-\frac{1}{4}Rg_{\mu\nu}\right)g_{\lambda\rho}+\frac{2\mathcal{C}l_{\text{P}}^2}{25\pi}T_{\mu\nu\lambda\rho}\right]n^{\mu}n^{\nu}n^{\lambda}n^{\rho}=0.
\end{equation}
In~\ref{n-dependence} we show that this equation implies vanishing of the fully symmetrised part of the tensor
\begin{equation}
\label{rank four}
\left(R_{(\mu\nu\vert}-\frac{1}{4}Rg_{(\mu\nu\vert}\right)g_{\vert\lambda\rho)}+\frac{2\mathcal{C}l_{\text{P}}^2}{25\pi}T_{\mu\nu\lambda\rho}=0.
\end{equation}
A rank $4$ fully symmetric tensor in principle contains $35$ independent components. However, all the terms in equation~\eqref{rank four} are of course constructed only from the auxiliary (since we work with a WTDiff-invariant theory) metric tensor and its derivatives, which only has $9$ independent components. To recover the $9$ relevant equations, we can simply take the trace of equation~\eqref{rank four}, obtaining
\begin{equation}
8\left(R_{\mu\nu}-\frac{1}{4}Rg_{\mu\nu}\right)+6\frac{2\mathcal{C}l_{\text{P}}^2}{25\pi}T_{\mu\nu\lambda}^{\;\:\;\:\;\:\lambda}=0.
\end{equation}
Since the Bel-Robinson tensor in $4$ spacetime dimensions is fully traceless, the correction term vanishes and we recover the traceless Einstein equations
\begin{equation}
R_{\mu\nu}-\frac{1}{4}Rg_{\mu\nu}=0.
\end{equation}
In other words, thermodynamic reasoning leads to no local vacuum corrections to gravitational dynamics at the order $O\left(l_{\text{P}}^2\right)$ if the local Lorentz invariance is preserved. This is consistent with the non-existence of any local, purely metric, diffeomorphism-invariant (or WTDiff-invariant) theory in $4$ spacetime dimensions whose action is at most quadratic in the Riemann tensor and whose equations of motion are not solved by the vacuum solutions of general relativity~\cite{Lovelock:1971} (or Weyl transverse gravity~\cite{Carballo:2022}).

\subsection{Vector-tensor interpretation}

An alternative interpretation of equation~\eqref{constraint} exists, which allows nontrivial corrections proportional to the Bel-Robinson tensor to survive. The price to be paid is losing the standard assumption of the local Lorentz invariance of physics and dealing with the challenges it poses for the self-consistency of our approach and its experimental and observational viability. Herein, we only briefly outline this alternative interpretation, leaving for a future analysis the conceptual questions it necessarily raises.

The key point is that the unit, timelike, future-pointing vector field $n^{\mu}$ is not arbitrary, but actually corresponds to some privileged direction of time. Such a preferred measure of time is introduced by some alternative theories of gravity, e.g. by Einstein-aether theory~\cite{Jacobson:2001} or by Ho\v{r}ava-Lifshitz gravity~\cite{Horava:2009}. Of course, giving $n^{\mu}$ a privileged status necessarily breaks the local Lorentz invariance, whose potential violations are very strongly bounded by experiments~\cite{Liberati:2013}. Therefore, any model violating this invariance faces the challenge of showing its compliance with these bounds. Nevertheless, the breaking of the local Lorentz invariance by introducing a preferred direction of time has been analysed, e.g. in the context of Ho\v{r}ava-Lifshitz gravity and found in principle experimentally viable~\cite{Wang:2017}.

Another challenge to this interpretation comes from the importance the local Lorentz invariance plays in the definition of causal horizons and, consequently, in their thermodynamic interpretation. However, it has been argued that black hole entropy survives in Einstein-aether and Ho\v{r}ava-Lifshitz gravity gravity~\cite{Berglund:2012,Cropp:2014}. Likewise, entanglement entropy associated with causal horizons has been studied in certain Lorentz invariance violating theories and found to obey the area law to the leading order~\cite{Solodukhin:2011}. Furthermore, it has been recently proposed that the Unruh effect occurs even if the local Lorentz invariance is broken~\cite{Liberati:2023} (and the same has been suggested for the Hawking effect~\cite{Cropp:2014}). In total, there appears to be a number of works arguing that the connection between gravity and thermodynamics survives beyond the local Lorentz invariance, perhaps allowing its use to derive the equations governing the gravitational dynamics even in that case (as already discussed in the context of momentum-dependent geometry~\cite{Chirco:2022}).

Taking $n^{\mu}$ as the preferred direction of time, the local equilibrium conditions lead to a single equation
\begin{equation}
\label{modified}
\left(R_{\mu\nu}-\frac{1}{4}Rg_{\mu\nu}\right)n^{\mu}n^{\nu}=-\frac{2\mathcal{C}l_{\text{P}}^2}{25\pi}\left(E^2+B^2\right),
\end{equation}
where we used equation~\eqref{EM Bel} for the electric-magnetic decomposition of the Bel-Robinson super-energy density. The approach we employ cannot straightforwardly recover any additional information. Therefore, to fully determine the gravitational dynamics in this way would require non-trivial modifications to our thermodynamic approach, which we leave for a future study.

For any spacetime with vanishing scalar curvature, $R=0$, the left hand side of equation~\eqref{constraint} can be written as
\begin{equation}
G_{\mu\nu}n^{\mu}n^{\nu}=-\frac{2\mathcal{C}l_{\text{P}}^2}{25\pi}\left(E^2+B^2\right),
\end{equation}
This equation has the form of the Hamiltonian constraint of general relativity with a non-trivial right hand side given by the Bel-Robinson super-energy density multiplied by a Planck scale area, which has the correct dimensions to be the quasilocal energy density of the gravitational field. Unfortunately, while suggestive, this interpretation of equation~\eqref{modified} cannot be directly applied due to the WTDiff-invariant nature of the gravitational dynamics implied by thermodynamics. The problem of the Hamiltonian constraint in Weyl transverse gravity (or any unimodular theory) is subtle~\cite{Kluson:2023,Kluson:2024,Herrera:2024}, since it cannot simply correspond to the time-time component of the traceless equations of motion which contains second time derivatives. Instead, equation~\eqref{modified} represents simply a generalisation of the time-time equation of motion of Weyl transverse gravity. Its interpretation in the Hamiltonian description of the theory requires a more careful analysis.

\subsection{Conceptual reflections on the possible Lorentz invariance violations}

The gravitational dynamics with the preferred direction of time we proposed in the previous subsection of course raises a number of possible objections. Herein, we provide a brief summary of these objections, as well as of the reasons to study this interpretation further.

The most notable objections are
\begin{itemize}
\item The local equilibrium conditions we studied directly imply only a single scalar equation. Clearly, determining the full dynamics of both the metric and the unit, timelike vector field $n^{\mu}$, requires additional equations. We are presently not aware of any straightforward way to derive them.
\item Both the local causal horizons and the Unruh effect implicitly rely on the local Lorentz invariance. However, both concepts have been studied in the literature in some local Lorentz invariance violating settings and found to be robust even in that case~\cite{Berglund:2012,Cropp:2014,Liberati:2023}. One would just need to carefully check their validity in the specific scenario we consider.
\item The experimental bounds on the violations of the Lorentz invariance are very stringent. Nevertheless, Ho\v{r}ava-Lifshitz gravity which violates the local Lorentz invariance in a way analogous to our proposal has been found to be in principle experimentally viable. It suggest that the approach we consider might be compatible with the experimental bounds as well.
\end{itemize}
To sum up, the local Lorentz invariance violating interpretation of our result remains highly speculative. Nevertheless, we find it interesting for several reasons
\begin{itemize}
\item The equation we find has a very natural form of the Ricci curvature being source by a candidate expression for a quasilocal energy of the gravitational field.
\item The Bel-Robinson term implies perturbative corrections to dynamics of vacuum black holes as well as to the gravitational wave propagation. Such corrections are phenomenologically interesting and might in principle make the theory experimentally falsifiable in the relatively near future (especially if they imply any changes to the speed of the gravitational waves~\cite{Zumalacarregui:2017}).
\item Depending on the sign of the correction term (essentially determined by the sign of the logarithmic correction to entropy), the Bel-Robinson term could allow the construction of various exotic solutions, such as wormholes, in vacuum.
\item Further analysis of the local Lorentz invariance violating interpretation of the local equilibrium conditions (especially going to the sixth order in the derivatives of the metric) might lead to a result more similar to Ho\v{r}ava-Lifshitz gravity, possibly even corresponding to some specific version of that theory. Given the status of Ho\v{r}ava-Lifshitz gravity as a possible UV completion of general relativity, obtaining it from thermodynamics of local causal horizons would be quite interesting.
\end{itemize}
To conclude, we find the potential benefits of the introduction of a preferred direction of time in the context of thermodynamics of local causal horizons sufficient to warrant its further study, despite the issues it raises.

\section{Discussion}
\label{discussion}

In this work, we examine whether local equilibrium conditions encode corrections to the traceless Einstein equations in vacuum. Previously, we have shown that perturbative corrections of this kind do occur in the presence of matter fields~\cite{Alonso:2023,Alonso:2020b,Alonso:2023b}. However, these works focused on extracting specific types of correction terms and neglected the possible contributions quadratic in Weyl tensor on which we focus here. Our main message is that, under the assumptions of locality and local Lorentz invariance, local equilibrium conditions imply no modifications to the equations governing gravitational dynamics that are perturbatively relevant in the vacuum at the order $O\left(l_{\text{P}}^2\right)$. In other words, the vacuum solutions to the Einstein equations acquire no corrections in this regime (although there might exist novel non-perturbative vacuum solutions).

The previous works have already shown how the area (and, hence, entropy) of local causal diamonds in vacuum changes under a small variation of geometry away from the flat spacetime one~\cite{Jacobson:2017,Wang:2019}. It has been established that the area variation is given by an expression quadratic in Weyl tensor. However, only for the light-cone cut local causal diamond and there only in $4$ spacetime dimensions, is the variation proportional to the Bel-Robinson super-energy density~\cite{Wang:2019} (at least without introducing additional rules for comparing the flat and the perturbed geometry~\cite{Jacobson:2017}). Even in that case, the variation of the area did not correctly encode the (traceless) Einstein equations in the presence of the matter fields~\cite{Wang:2019}, making it impossible to employ it to study corrections to gravitational dynamics.

Herein, we also applied the light-cone cut local causal diamond construction, but with three key differences compared to the previous analysis~\cite{Wang:2019}.
\begin{itemize}
\item We track the change of the area along the null geodesic generators of the horizon in a generic spacetime rather than considering its change due to a small variation of the geometry away from flat spacetime. This choice allows us to sidestep fixing the (generalised) volume of the spatial slices of the causal diamonds~\cite{Jacobson:2017,Wang:2019}.
\item We set the initial conditions for the evolution of expansion and shear so that both vanish at the bifurcate surface of the horizon (at $\lambda=0$), rather than at the past apex ($\lambda=-l$) as in the previous work. While the precise initial condition for the shear could probably be left unfixed~\cite{Chirco:2010} (we plan to address this issue in a future work), the one for expansion has important consequences. It is easy to check that setting expansion to zero in the past apex leads to an infinite jump of its value at $\lambda=-l$, which vanishing expansion at the bifurcate surface avoids. Furthermore, our choice causes the expansion to change its sign at the bifurcate surface as required (since, by definition, the diamond stops expanding and starts contracting there). Consequently, our initial conditions allow us to derive the traceless Einstein equations in the presence of matter, which the previously adapted choice made impossible.
\item We consider the logarithmic term in entropy of local causal horizons. The leading order term proportional to area is $O\left(1/l_{\text{P}}^2\right)$, whereas the logarithmic correction is $O\left(l_{\text{P}}^0\right)$. Thence, including this term bring a new length scale in the form of the Planck length, which makes the presence of local corrections to the Einstein equations possible (since such corrections have necessarily different length dimension than the Ricci tensor).
\end{itemize}
As we stressed, this procedure leads to no locally Lorentz invariant modifications to gravitational dynamics in vacuum. However, we have previously shown, albeit in a slightly different geometric setup, that it implies nontrivial corrections to the traceless Einstein equations in the presence of matter~\cite{Alonso:2020b}. The techniques developed in this paper should allow us to eventually address this case in full generality and derive all the relevant correction terms at the order $O\left(l_{\text{P}}^2\right)$.

We also proposed an alternative, local Lorentz invariance violating, interpretation, which does imply modified gravitational dynamics. The result has the form of the Bel-Robinson super-energy density multiplied by a Planck scale area acting as a source for the time-time component of the traceless Ricci tensor. This outcome agrees remarkably well with the interpretation of the Bel-Robinson super-energy density as the quasilocal gravitational energy density per unit area. We stress that this alternative approach faces a number of challenges, most notably in recovering its the thermodynamic assumptions without the local Lorentz invariance. Moreover, we have been able to only outline it, as we presently lack the remaining equations of motion. Despite its shortcomings, this approach also has a number of attractive features. Due to the natural appearance of the Bel-Robinson tensor, it implies perturbative modifications to vacuum gravitational dynamics. Furthermore, it offers the possibility of making contact with the power-counting renormalisable Ho\v{r}ava-Lifshitz gravity. In conclusion, we believe that the Lorentz invariance violating interpretation of our results deserves further attention.

\section*{Acknowledgements}

The authors thank Luis J. Garay for helpful discussions of issues closely related with this works. AA-S is funded by the Deutsche Forschungsgemeinschaft (DFG, German Research Foundation) — Project ID 51673086. ML is supported by the DIAS Post-Doctoral Scholarship in Theoretical Physics 2024 and by the Charles University Grant Agency project No. GAUK 90123. AA-S acknowledges support through Grants No. PID2020–118159 GB-C44 and PID2023-149018NB-C44 (funded by MCIN/AEI/10.13039/501100011033).

\appendix

\section{Removing contractions with an arbitrary timelike vector}
\label{n-dependence}

In this appendix, we show that if a rank $k$ tensor $Q_{\alpha_1\dots\alpha_k}$ obeys $Q_{\alpha_1\dots\alpha_k}n^{\alpha_1}\dots n^{\alpha_k}=0$ for an arbitrary unit, timelike, future-pointing vector $n^{\mu}$, its fully symmetric part must vanish. For this purpose, we introduce an orthonormal basis $e^{0}$, $e^{1}$,\dots,$e^{D-1}$. In this basis, any unit, timelike, future-pointing vector $n^{\mu}$ can be decomposed as
\begin{equation}
n^{\mu}=\sqrt{1+\sum_{i=1}^{D-1}p_i^2}+\sum_{i=1}^{D-1}p_ie^i,
\end{equation}
where $p_i$ are real numbers. Condition  $Q_{\alpha_1\dots\alpha_k}n^{\alpha_1}\dots n^{\alpha_k}=0$ must then hold for arbitrary values of $p_i$. Any component of $Q_{\alpha_1\dots\alpha_k}$, up to permutations of indices, is then multiplied by a unique combination of powers of various $p_i$'s and of the factor $\sqrt{1+\sum_{i=1}^{D-1}p_i^2}$. Therefore, every component of the totally symmetrised tensor $Q_{(\alpha_1\dots\alpha_k)}$ must be separately equal to zero and $Q_{(\alpha_1\dots\alpha_k)}$ indeed vanishes.

\section{Non-local results for general spacetime dimensions}

In the main text, we strictly assumed the equations governing gravitational dynamics to be local. Then, these equations cannot depend on an arbitrary diamond size parameter $l$. At this point, we cannot physically motivate any scenario where $l$ does affect the gravitational dynamics. Nevertheless, the $l$-dependent results are easy to obtain and rather interesting in their own right. Hence, we briefly discuss them in this appendix. Let us then assume that $l$ either takes a preferred value (corresponding to some non-locality scale), or that our choice of $l$ somehow affects the gravitational dynamics. We still require the condition $l\gg l_{\text{P}}$ to view the spacetime as a smooth manifold. Consequently, the $l$-dependent corrections to the equations become much larger than the corrections coming from the logarithmic contribution to entropy, which are suppressed by $l_{\text{P}}^2$, and we can work with a simple vacuum equilibrium condition $\Delta\mathcal{A}_{\text{rev}}=0$. Since we do not rely on the logarithmic term, which is the leading order correction to entropy only in $4$ spacetime dimensions, we can carry out the derivation in an arbitrary dimension $D$. In this case, we find for $\Delta\mathcal{A}_{\text{rev}}$
\begin{align}
\nonumber &\Delta\mathcal{A}_{\text{rev}}=-\frac{\Omega_{D-2}l^{D}}{\left(D-1\right)^2}\left(R_{\mu\nu}-\frac{1}{D}Rg_{\mu\nu}\right)n^{\mu}n^{\nu}-\frac{6\Omega_{D-2}l^{D+2}n^{\mu}n^{\nu}n^{\lambda}n^{\rho}}{\left(D-1\right)^2D\left(D+1\right)^2\left(D+2\right)} \\
&\left[\left(D+2\right)\left(D+4\right)C_{(\mu\vert\sigma\vert\nu\vert\tau}C_{\vert\lambda\;\:\rho)}^{\;\:\sigma\;\:\tau}
-3\left(D+2\right)C_{(\mu\vert\sigma\alpha\beta}C_{\vert\nu}^{\;\:\sigma\alpha\beta}g_{\lambda\rho)}+\frac{3}{2}g_{\mu\nu}g_{\lambda\rho}C_{\sigma\alpha\beta\gamma}C^{\sigma\alpha\beta\gamma}\right].
\end{align}
The symmetric tensor on the second line is completely traceless. As noted in an earlier work~\cite{Wang:2019}, it only coincides with the Bel-Robinson tensor in $4$ dimensions. Otherwise, it might be thought of as the traceless part of the Bel-Robinson tensor. This result makes sense if we look at it as a generalisation of Weyl transverse gravity, whose source term is likewise just the traceless part of the energy-momentum tensor.

Mathematically, the tracelessness of the expression is a simple consequence of evaluating the change in area between two cuts of a null surface (as opposed to a variation of a spacelike geodesic ball~\cite{Jacobson:2017}). Indeed, the integral of the expression $k^{\mu}k^{\nu}k^{\lambda}k^{\rho}$ over a unit sphere yields
\begin{align}
\nonumber \int k^{\mu}k^{\nu}k^{\lambda}k^{\rho}\text{d}\Omega_{D-2}=&\frac{\left(D+2\right)\left(D+4\right)}{\left(D-1\right)\left(D+1\right)}n^{\mu}n^{\nu}n^{\lambda}n^{\rho}+\frac{24\left(D+2\right)}{\left(D-1\right)\left(D+1\right)}g^{(\mu\nu}n^{\lambda}n^{\rho)} \\
&+\frac{24}{\left(D-1\right)\left(D+1\right)}g^{(\mu\nu}g^{\lambda\rho)},
\end{align}
which is a completely traceless expression.

Assuming the local Lorentz invariance, we again recover the traceless Einstein equations without any corrections (due to tracelessness). Allowing for a preferred direction of time, we obtain the following scalar equation
\begin{align}
\nonumber &\left(R_{\mu\nu}-\frac{1}{D}Rg_{\mu\nu}\right)n^{\mu}n^{\nu}=-\frac{6\left(D+2\right)\left(D+4\right)l^2n^{\mu}n^{\nu}n^{\lambda}n^{\rho}}{D\left(D+1\right)^2\left(D+2\right)} \\
&\left[\left(D+2\right)\left(D+4\right)C_{(\mu\vert\sigma\vert\nu\vert\tau}C_{\vert\lambda\;\:\rho)}^{\;\:\sigma\;\:\tau}
-3\left(D+2\right)C_{(\mu\vert\sigma\alpha\beta}C_{\vert\nu}^{\;\:\sigma\alpha\beta}g_{\lambda\rho)}+\frac{3}{2}g_{\mu\nu}g_{\lambda\rho}C_{\sigma\alpha\beta\gamma}C^{\sigma\alpha\beta\gamma}\right].
\end{align}
As in $4$ dimensions, this equation breaks the local Lorentz invariance. However, now it also contains a non-local parameter $l$.

\section*{References}
\bibliography{bibliography}
\bibliographystyle{iopart-num}

\end{document}